\documentclass[10pt,conference,letterpaper]{IEEEtran}
\IEEEoverridecommandlockouts 
\usepackage{enumerate, verbatim, enumitem}
\usepackage{graphicx, float}
\usepackage{amsmath, amssymb, amsthm, mathtools, stmaryrd}
\usepackage{cleveref, xcolor, todonotes, cite}
\usepackage{array, tabularx, booktabs}
\usepackage{mdframed}

\def\BibTeX{{\rm B\kern-.05em{\sc i\kern-.025em b}\kern-.08em
    T\kern-.1667em\lower.7ex\hbox{E}\kern-.125emX}}

\newcommand{\pad}{\addvspace{2ex}}
\newcommand{\proofover}{\hfill\IEEEQED\par\addvspace{1.5ex}}

\newcommand{\mlo}{\mu_{l1}}
\newcommand{\mlt}{\mu_{l2}}
\newcommand{\mco}{\mu_{c1}}
\newcommand{\mct}{\mu_{c2}}

\newcommand{\lp}{\left(}
\newcommand{\rp}{\right)}
\newcommand{\lsb}{\left[}
\newcommand{\rsb}{\right]}
\newcommand{\lcb}{\left\{}
\newcommand{\rcb}{\right\}}

\newcommand{\Pl}{\mathcal P_\lambda}
\newcommand{\Fl}{\mathcal F_\lambda}
\newcommand{\tssm}{T^*_{\text{TM}}}
\newcommand{\lsm}{\Lambda_{\text{TM}}}
\newcommand{\ldm}{\Lambda_{\text{DM}}}
\newcommand{\tsm}{T_{\text{TM}}}
\newcommand{\tdm}{T_{\text{DM}}}
\newcommand{\toh}{T_{\text{OH}}}

\newcommand{\given}{\,|\,} 

\theoremstyle{remark}
\newtheorem*{remark}{Remark}

\newmdtheoremenv[
  linewidth=0.8pt,
  topline=true,
  bottomline=true,
  leftline=true,
  rightline=true,
  linecolor=black,
  backgroundcolor=white, 
  innertopmargin=6pt,
  innerbottommargin=6pt,
  innerleftmargin=6pt,
  innerrightmargin=6pt,
  skipabove=10pt,
  skipbelow=10pt,
]{lemma}{\upshape{\textbf{Lemma}}}

\newmdtheoremenv[
  linewidth=0pt,
  topline=true,
  bottomline=true,
  leftline=true,
  rightline=true,
  linecolor=black,
  backgroundcolor=white, 
  innertopmargin=6pt,
  innerbottommargin=6pt,
  innerleftmargin=6pt,
  innerrightmargin=6pt,
  skipabove=10pt,
  skipbelow=10pt,
]{claim}{\upshape{\textbf{Claim}}}

\newmdtheoremenv[
  linewidth=0.8pt,
  topline=true,
  bottomline=true,
  leftline=true,
  rightline=true,
  linecolor=black,
  backgroundcolor=white, 
  innertopmargin=6pt,
  innerbottommargin=6pt,
  innerleftmargin=6pt,
  innerrightmargin=6pt,
  skipabove=10pt,
  skipbelow=10pt,
]{proposition}{\upshape{\textbf{Proposition}}}

\newmdtheoremenv[
  linewidth=0.8pt,
  topline=true,
  bottomline=true,
  leftline=true,
  rightline=true,
  linecolor=black,
  backgroundcolor=white, 
  innertopmargin=6pt,
  innerbottommargin=6pt,
  innerleftmargin=6pt,
  innerrightmargin=6pt,
  skipabove=10pt,
  skipbelow=10pt,
]{corollary}{\upshape{\textbf{Corollary}}}

\newmdtheoremenv[
  linewidth=1pt,
  topline=true,
  bottomline=true,
  leftline=true,
  rightline=true,
  linecolor=black,
  backgroundcolor=white, 
  innertopmargin=6pt,
  innerbottommargin=6pt,
  innerleftmargin=6pt,
  innerrightmargin=6pt,
  skipabove=10pt,
  skipbelow=10pt,
]{theorem}{\upshape{\textbf{Theorem}}}

\mdtheorem[
  linecolor=black,
  linewidth=0.8pt,
  topline=true,
  bottomline=true,
  leftline=true,
  rightline=true,
  backgroundcolor=white,
  innertopmargin=6pt,
  innerbottommargin=6pt,
  innerleftmargin=4pt,
  innerrightmargin=5pt,
  skipabove=10pt,
  skipbelow=10pt,
  leftmargin=0pt,
  rightmargin=0pt,
  frametitlefont=\normalfont\bfseries,
  frametitlealignment=\raggedright,
  frametitleaboveskip=3pt,       
  frametitlebelowskip=-1pt,      
  font=\normalfont,
]{definition}{Definition}

\begin{document} 
\title{Delay Optimization in a Simple Offloading System: Extended Version}

\author{
\IEEEauthorblockN{Darin Jeff and Eytan Modiano}
\IEEEauthorblockA{LIDS, Massachusetts Institute of Technology, Cambridge, MA, USA\\
Email: \{djeff, modiano\}@mit.edu}
}

\maketitle

\begin{abstract}
We consider a computation offloading system where jobs are processed sequentially at a local server followed by a higher-capacity cloud server. The system offers two service modes, differing in how the processing is split between the servers. Our goal is to design an optimal policy for assigning jobs to service modes and partitioning server resources in order to minimize delay. We begin by characterizing the system's stability region and establishing design principles for service modes that maximize throughput. For any given job assignment strategy, we derive the optimal resource partitioning and present a closed-form expression for the resulting delay. Moreover, we establish that the delay-optimal assignment policy exhibits a distinct breakaway structure: at low system loads, it is optimal to route all jobs through a single service mode, whereas beyond a critical load threshold, jobs must be assigned across both modes. We conclude by validating these theoretical insights through numerical evaluation.
\end{abstract}




\section{Introduction}\label{sec:introduction}
The rapid adoption of Large Language Models (LLMs) such as GPT-4, Claude, and Mistral has significantly increased demand for cloud computing resources. These models require memory-intensive architectures that generally exceed the capabilities of consumer-grade hardware, leading most LLM-based applications to rely heavily on a centralized cloud infrastructure for inference and fine-tuning.

At the same time, end-user devices are becoming increasingly capable. Hardware manufacturers such as Apple, AMD, and Qualcomm have recently announced plans for consumer-grade devices that can run models with a few billion parameters locally. In parallel, a growing body of work \cite{LLM1,LLM2,LLM4} has explored strategies for partitioning LLM workloads across user and cloud devices. Together, these trends motivate a re-examination of conventional cloud-centric processing pipelines. Instead of the conventional strategy of offloading most of the computation to the cloud, systems can adopt alternative strategies that place a greater portion of the workload on user devices when appropriate, thereby improving overall system capacity and delay performance.

Recently, there has been a tremendous amount of interest in the topic of computational offloading.
Qin et al.\cite{qin2023efficient, qin2023distributed} proposed distributed threshold-based offloading policies for mobile cloud computing systems, formulating queue-aware decision-making as a game-theoretic problem and establishing equilibrium properties. Similarly, Zhou et al.\cite{zhou2021distributed} utilized queue-length-based thresholds in multi-agent Mobile Edge Computing systems and demonstrated that their distributed best-response algorithm achieves near-optimal offloading utility compared to centralized benchmarks. Lyapunov-based control techniques have also been explored in distributed computing networks~\cite{Hao, Andreas}, where drift-plus-penalty methods are used to design algorithms that are throughput-optimal while enabling tunable trade-offs between delay and cost in offloading settings. Finally, in closely related settings, the authors in \cite{jeff} showed that delay-optimal queue-length-based policies exhibit a switch-type structure.

While prior work has focused on optimizing computation offloading in specific system models, our goal is to develop a more fundamental understanding of such systems. To that end, we study a two-stage offloading system in which each job is processed sequentially by a local server with limited capacity, followed by a more powerful cloud server. Jobs are served using one of two service modes, each specifying how the job’s workload is partitioned across the two servers. One mode is "cloud-heavy", offloading a larger fraction of computation, while the other is "local-heavy" and relies more heavily on local processing. This dual-mode structure captures systems that have traditionally favored cloud-centric execution but are increasingly capable of greater local processing due to advances in end-user hardware.

\begin{remark}[Model generality]
Although our model is motivated by computational offloading, it is not inherently tied to computation. The two-stage structure also captures settings in which limited communication capacity, rather than processing power, is the primary system bottleneck. In such cases, the first stage can be interpreted as local preprocessing (e.g., compression or feature extraction), while the second stage represents transmission over a constrained communication link. The analysis and structural insights developed in this paper apply equally to these settings.
\end{remark}

The primary goal of this paper is to characterize the optimal mode-assignment and server resource allocation strategy that stabilizes the system for all feasible arrival rates while minimizing average job delay.

We first characterize the stability region of the proposed dual-mode system and highlight key system design considerations, showing how poorly chosen service modes can introduce throughput bottlenecks. We then identify fundamental trade-offs between delay performance and achievable throughput. Finally, in the general case, we establish that the delay-optimal operating policy exhibits a simple yet intuitive breakaway structure: at low arrival rates, it is optimal to exclusively use the "cloud-heavy" mode, whereas beyond a critical threshold arrival rate, assigning some jobs to "local heavy" processing becomes necessary.                                                                                                                                                                                                                                                                                                                                                                                                                                                                                                                                                                                                                                                                                                                                                                                                                                                                                                                                                                                                                                                                                                                                                                                                                                                                                                                                                 

\medskip
The remainder of the paper is organized as follows. Section~\ref{sec:system_model} introduces the dual-mode system model. Section~\ref{sec:prelim_anal} develops preliminary analytical tools and results that underpin the subsequent analysis. In Section~\ref{sec:stability_analysis}, we characterize the system’s stability region and discuss key design implications. Section~\ref{sec:optimal_resource_allocation} derives the optimal allocation of server resources for a given mode assignment strategy. Section~\ref{sec:optimal_policy} establishes structural properties of the optimal assignment policy. Section~\ref{sec:simulations} presents simulation results that validate the analytical findings. Finally, Section~\ref{sec:conclusion} concludes the paper.

\section{System Model}\label{sec:system_model}
In this section, we present the system model, which we call the \textit{dual-mode} system. The system consists of two sequential servers - a local server followed by a cloud server. Incoming jobs are processed using one of two available service modes (SM). Each server maintains separate queues and partitions its computing resources to serve these modes independently.
Jobs arrive following a Poisson process with rate $\lambda$. Upon arrival, each job is assigned to one of the two service modes: SM1 with probability $p$, or SM2 with probability $\bar{p} = 1 - p$. Each server reserves a fixed portion of its resources exclusively for each mode. Specifically, the fractions $\alpha$ and $\beta$ represent the local and cloud resources allocated to SM1, with the complementary fractions $\bar\alpha$ and $\bar\beta$ allocated to SM2. Thus, the system's \textit{operating point} is defined by the \textit{assignment parameter} $p$ and the \textit{partition parameters} $\alpha$ and $\beta$.

\begin{figure}[!b]
\centering
\includegraphics[width=0.4\textwidth]{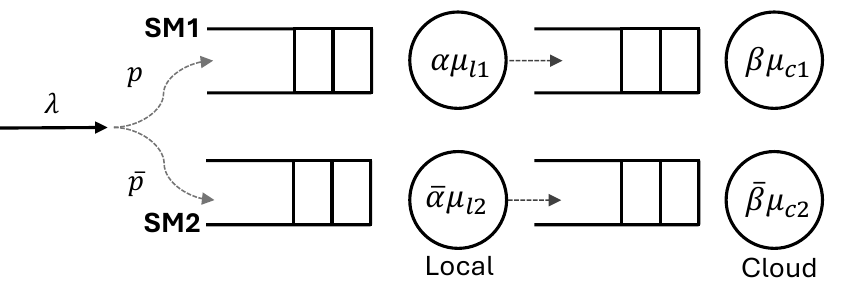}
\caption{Dual-Mode System depicting job arrivals, probabilistic service mode assignment, and dedicated resource allocation at the servers.}
\label{fig:dualmode}
\end{figure}


We denote by $\mlo$ and $\mco$ the service rates at the local and cloud servers, respectively, when their entire capacities are dedicated to SM1 jobs. Similarly, $\mlt$ and $\mct$ represent full-capacity service rates for SM2. Without loss of generality, we designate SM1 as the "cloud-heavy" mode, relying more heavily on cloud processing. Correspondingly, SM2 is the "local-heavy" mode, with a greater share of computation performed at the local server. Formally, this is represented by:
\[\mco < \mct \quad \text{and} \quad \mlt < \mlo\]

At a given operating point $(p, \alpha, \beta)$, SM1 jobs experience independent, exponentially distributed service times with rates $\alpha \mlo$ and $\beta \mco$ at the local and cloud servers, respectively. Similarly, SM2 jobs are served at rates $\bar{\alpha} \mlt$ and $\bar{\beta} \mct$. This is illustrated in Figure~\ref{fig:dualmode}.

The objective is to determine the optimal operating point that minimizes the average delay experienced by jobs.

\subsection{Canonical Transformation}
Up to this point, we have described the system using explicit service rates associated with each mode. While directly observable, this parameterization can obscure underlying structural insights. To address this, we introduce a canonical transformation that maps the rate parameters into parameters that capture (i) the effective capacity of the local server, (ii) the relative capacity of the cloud server compared to the local server, and (iii) the division of workload across service modes.

Define 
\begin{align*}
\mu_0 &\coloneqq \frac{\mlo\mlt(\mct-\mco)}{\mlo\mct - \mlt\mco} 
&& K \coloneqq \frac{\mco\mct(\mlo - \mlt)}{\mlo\mlt(\mct - \mco)} \nonumber \\
f_1 &\coloneqq \frac{\mlt(\mct-\mco)}{\mlo\mct - \mlt\mco} 
&& f_2 \coloneqq \frac{\mlo(\mct-\mco)}{\mlo\mct - \mlt\mco} 
\end{align*}

Here, $\mu_0$ and $K\mu_0$ capture the effective processing capacities of the local and cloud servers, respectively. The parameters $f_1$ and $f_2$ represent the fraction of total job workload executed locally in service modes SM1 and SM2. Under this interpretation, a job's workload is divided between local and cloud processing according to the ratio $f_i : (1 - f_i)$ for mode $i$.
The original service rates can be uniquely recovered from this representation using the following mapping:


\begin{align*}
    \mlo &= \frac{\mu_0}{f_1}, &
    \mlt &= \frac{\mu_0}{f_2}, &
    \mco &= \frac{K\mu_0}{1 - f_1}, &
    \mct &= \frac{K\mu_0}{1 - f_2} 
\end{align*}

Our analysis later shows that the order of servers does not affect the system's stability region or delay performance. Thus, without loss of generality, we take:\[K>1\]indicating that the cloud server has greater processing capacity compared to the local server. In the following section, we conduct preliminary analysis and develop essential analytical tools to study this system.

\section{Preliminary Analysis}\label{sec:prelim_anal}
In the previous section, we introduced a canonical representation for the dual-mode system that explicitly captures server capacities and workload distribution. Extending this perspective - where a service mode defines how job load is distributed across servers - we now consider an offloading system with a single, adjustable service mode, which we call the \textit{tunable-mode system}. We later demonstrate that this simpler system provides a useful benchmark for evaluating the stability and delay performance of the dual-mode system with comparable server capacities.

\subsection{Tunable-Mode System}
The tunable-mode system comprises two sequential servers: a local server with capacity $\mu_0$ and a cloud server with higher capacity $K\mu_0$, where $K>1$. Jobs arrive according to a Poisson process with rate $\lambda$.

\begin{figure}[!b]
\centering
\includegraphics[width=0.4\textwidth]{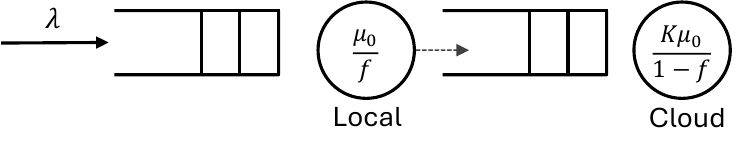}
\caption{Tunable-Mode System depicting job arrivals, and service times at each server.}
\label{fig:singlemode}
\end{figure}

Each incoming job is processed under the same service strategy, characterized by a tunable parameter $f \in [0,1]$, which determines the fraction of processing allocated to the local server. This parameter is fixed during system operation, but can be optimized offline based on system characteristics. Each server maintains a queue for incoming tasks, as depicted in Figure~\ref{fig:singlemode}.

Under \textit{service fraction parameter} $f$, the processing times at each server follow independent exponential distributions with service rates $\mu_0/f$ at the local server and $K\mu_0/(1-f)$ at the cloud server. This ensures a constant expected total processing requirement per job, independent of $f$.

The system objective is to select the optimal service fraction parameter, $f\in[0,1]$ under which the average delay in the system is minimized.

\subsection{Delay Performance}
The single-mode configuration evolves as a standard two-node Jackson tandem network, which is well-established in classical queueing theory (see, e.g.,~\cite{datanetworks}). Classical results establish that stability is impossible if the arrival rate $\lambda$ equals or exceeds the combined system capacity $\mu^*$, defined as:
\[\mu^* = (K+1)\mu_0.\]

We define the stability region $\lsm$ (mnemonically tunable-mode) as:
\[\lsm = \{\lambda > 0 \given \lambda < \mu^*\}\]
For arrival rates $\lambda \in \lsm$, the system remains stable if the \textit{service-fraction} parameter $f$ lies within the set $\Fl$, defined as:
\[\Fl \coloneqq [0,1] \cap \lp1-\dfrac{K\mu_0}{\lambda}, \dfrac{\mu_0}{\lambda}\rp.\]
The expected delay in this system is given by:
\[\tsm(f;\lambda) \coloneqq \frac{f}{\mu_0 - f\lambda} + \frac{\bar{f}}{K\mu_0 - \bar{f}\lambda}\]
which captures both the wait and processing times.

We characterize the optimal service fraction parameter in the following theorem.

\begin{theorem}\label{th:tunable_mode}
For arrival rates $\lambda \in \lsm$, the delay-optimal service fraction parameter is:
\[f^*(\lambda) = \max\{f_{\min}(\lambda), 0\}\]
where 
\[f_{\min}(\lambda) \coloneqq \dfrac{\lambda - \mu_0(K-\sqrt K)}{\lambda (1+\sqrt K)}\] 
The corresponding average delay under $f^*(\lambda)$ is:
\begin{align*}
\tssm(\lambda) &\coloneqq \tsm(f^*(\lambda);\lambda)\\
&= \begin{cases}
\frac{2\lambda - (\sqrt K - 1)^2\mu_0}{\lambda \lp (K+1)\mu_0 - \lambda\rp}, & \lambda > (K-\sqrt{K})\mu_0 \\
\frac{1}{K\mu_0 - \lambda}, & \lambda \leq (K-\sqrt{K})\mu_0
\end{cases}
\end{align*}
\end{theorem}
\begin{IEEEproof}
For a fixed $\lambda\in\lsm$, the delay function $\tsm(\cdot;\lambda)$ is strictly convex on the interval $\left(1-\frac{K\mu_0}{\lambda}, \frac{\mu_0}{\lambda}\right)$, with a unique minimizer:
\[f_{\min}(\lambda) = \dfrac{\lambda - \mu_0(K-\sqrt K)}{\lambda (1+\sqrt K)}\]
Since $K>1$, $f_{\min}(\lambda)$ lies in the interval $[-\infty,\frac{1}{K+1})$ for $\lambda\in\lsm$. However, this value may fall outside the feasible range $[0,1]$. Due to strict convexity, the optimal feasible solution is either $f_{\min}(\lambda)$ if within $[0,1]$, or the nearest boundary point ($f=0$). The stated expressions follow directly from algebraic simplifications.
\end{IEEEproof}

Theorem~\ref{th:tunable_mode} shows that at lower arrival rates ($\lambda \leq (K-\sqrt{K})\mu_0$), the optimal service strategy assigns all processing exclusively to the cloud server (setting $f=0$). As the arrival rate increases beyond this threshold, it becomes beneficial to distribute the workload between local and cloud servers. Moreover, the optimal fraction allocated locally, $f^*(\lambda)$ always remains below $1/(K+1)$ -- which is the relative capacity of the local server to the total system capacity.

\section{Stability Analysis}\label{sec:stability_analysis}
In this section, we return to the original system model, leveraging results from our previous analysis. We analyze the stability region in terms of the original rate parameters and glean interpretive insights later using their canonical forms.

As described in Section~\ref{sec:system_model}, we represent service rates under the two service modes by $\mlo, \mlt, \mco, \mct$, and their corresponding canonical representation by $\mu_0, K, f_1, f_2$.

Although our analysis focuses on stationary randomized policies, classical network optimization literature \cite{networks} establishes that these policies fully characterize the stability region, even when dynamic, state-dependent policies are considered.

The characterization of the stability region naturally consists of two parts: an achievability argument, where we construct a stabilizing policy that supports all arrival rates within the region, and a converse, which shows that no policy can stabilize the system for rates outside this region. The stability region for the dual-mode system is formally stated below.

\begin{theorem}[Stability Region] \label{th:stability_region}
The stability region $\ldm$ for the dual-mode system is:
\[\ldm \coloneqq \{\lambda \geq 0: \lambda < \lambda_{\max}\}\]
where
\[\lambda_{\max} \coloneqq \min\{\mlo, \mct, \mu^*\}\] and \[\mu^* \coloneqq \frac{\mco\mct(\mlo-\mlt)+\mlo\mlt(\mct-\mco)}{\mlo\mct - \mlt\mco}\]
\end{theorem}
\begin{IEEEproof}
The result follows directly from Lemma~\ref{lem:acheivability} (achievability) and Lemma~\ref{lem:converse} (converse), which we state and prove next.	
\end{IEEEproof}

\begin{figure}[!b]
    \centering
    \begin{minipage}{0.8\linewidth}
        \includegraphics[width=\linewidth]{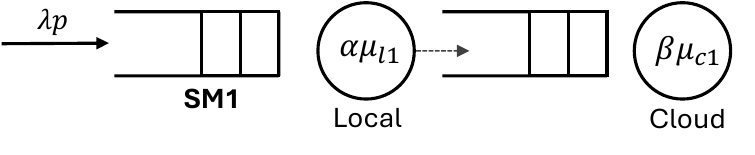}
        \smallskip\centering
    \small (a) Service Mode~1 subsystem
    \bigskip
    \end{minipage}
    \begin{minipage}{0.8\linewidth}
        \includegraphics[width=\linewidth]{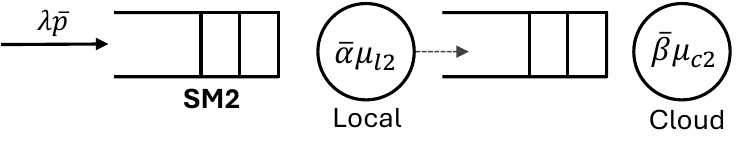}
    \smallskip\centering
    \small (b) Service Mode~2 subsystem
    \end{minipage}
    \caption{Independent subsystems resulting from Poisson splitting in the dual-mode configuration.}
    \label{fig:subsystems}
\end{figure}

Recall that in this system, incoming jobs are randomly divided into two subsystems associated with each service mode. Due to the Poisson splitting property, each subsystem behaves as an independent two-node Jackson tandem network with server capacities determined by parameters $\alpha$ and $\beta$, and arrival rate governed by parameter $p$, as illustrated in Figure~\ref{fig:subsystems}.

From standard stability criteria in queueing theory, the dual-mode system is stable if and only if the arrival rates into each server are strictly less than their corresponding service rates. Formally, the system is stable if and only if there exist parameters $p, \alpha, \beta \in [0,1]$ satisfying:
\begin{align}
\lambda p &< \alpha\mlo, \quad \lambda \bar p < \bar\alpha\mlt, \quad \lambda p < \beta\mco, \quad \lambda \bar p < \bar\beta\mct \label{eqn:stability_equations}
\end{align}


\begin{lemma}[Achievability]\label{lem:acheivability}
	For the dual-mode system, if 
	$\lambda \in \ldm$, then there exist $p,\alpha,\beta\in[0,1]$ such that the inequalities in \eqref{eqn:stability_equations} are satisfied.
\end{lemma}
\begin{IEEEproof}
The result follows by explicitly constructing values of $p$, $\alpha$, and $\beta$ that satisfy the inequalities in \eqref{eqn:stability_equations}. A detailed proof can be found in Appendix~\ref{app:stability_region}.
\end{IEEEproof}


\begin{lemma}[Converse]\label{lem:converse}
	For the dual-mode system, if \mbox{$\lambda \notin \ldm$}, then, the inequalities in \eqref{eqn:stability_equations} are not satisfied for any $p, \alpha, \beta \in [0,1]$.
\end{lemma}
\begin{IEEEproof}
The result follows via a contradiction argument. A detailed proof can be found in Appendix~\ref{app:stability_region}.
\end{IEEEproof}


\subsection{Interpretation}
We offer some intuitive reasoning to clarify the structure of this stability region characterized by:
\[\lambda_{\max} = \min\{\mlo, \mct, \mu^*\}\]

By convention, we have $\mlo>\mlt$ and $\mct>\mco$. Thus, $\mlo$ represents the maximum achievable processing rate at the local server, and $\mct$ represents the maximum achievable processing rate at the cloud server. To interpret $\mu^*$, we express it in terms of the canonical parameters:
\[\mu^* = (K+1)\mu_0\]
Recall, this was the same definition for $\mu^*$ in the tunable-mode system and represents the combined capacity of the local and cloud server.


\subsection{System Design Considerations}
In the single-mode configuration, we saw that a well-designed service mode can stabilize arrival rates up to $\lambda_{\max} = \mu^*$. In contrast, if $\lambda_{\max} < \mu^*$ in the dual-mode system, it indicates poor service mode design, with both modes heavily relying on either the local or cloud server, creating a bottleneck. Conversely, $\lambda_{\max}=\mu^*$ indicates a balanced design capable of fully utilizing both server capacities.

Formally, if we consider a system with $\mlo<\mu^*$, rewriting this condition in terms of the canonical parameters yields
\[\mlo<\mu^* \Leftrightarrow \frac{1}{K+1} < f_1 < f_2\]

Having both $f_1, f_2 > \frac{1}{K+1}$ means each service mode assigns workloads to the local server that exceed its proportional capacity $\mu_0$ relative to the overall system capacity $(K+1)\mu_0$ causing a throughput bottleneck.
Similarly, the case $\mct<\mu^*$ gives:
\begin{align*}
\mct<\mu^* \Leftrightarrow f_1 < f_2 < \frac{1}{K+1}
\end{align*}
which corresponds to systems inadequately utilizing the cloud's capacity due to excessive local processing.

Since maximizing the stability region is an important consideration in designing service modes for offloading systems, we define systems achieving this as \textit{throughput-efficient}:

\begin{definition} \label{def:throughput_efficient}
A dual-mode system is \textit{throughput efficient}
if its canonical service mode parameters satisfy:
\[f_1 < \frac{1}{K+1} < f_2.\]
\end{definition}

\section{Optimal Resource Allocation}\label{sec:optimal_resource_allocation}
In the previous section, we established the existence of stabilizing operating points for arrival rates within the stability region. Here, we characterize the delay-optimal partition parameters -- $\alpha^*, \beta^*$ for a given value of the assignment parameter -- $p$. For this, we need to ensure that stabilizing partition parameters exist for this $p$.
Following the stability analysis in Section~\ref{sec:stability_analysis}, we see that stabilizing choices of partition parameters exist for arrival rate $\lambda \in \ldm$, when the assignment parameter satisfies: 
\[p\in \mathcal \Pl \coloneqq [0,1]\cap(p_{\min},p_{\max})\]
where: $p_{\min} = 1 - \frac{1/\lambda-1/\mlo}{1/\mlt-1/\mlo}$ and $p_{\max} = \frac{1/\lambda-1/\mct}{1/\mco-1/\mct}$.\\

From standard queueing theory, the average delay in a dual-mode system operating stably at rate $\lambda \in \ldm$ is given by:
\begin{align*}
\tdm(p,\alpha,\beta; \lambda)
= &\frac{p}{\alpha\mlo-\lambda p} + \frac{\bar{p}}{\bar{\alpha}\mlt-\lambda \bar{p}} \\
 &+ \frac{p}{\beta\mco-\lambda p} + \frac{\bar{p}}{\bar{\beta}\mct-\lambda \bar{p}}
\end{align*}
We optimize this expression to obtain the optimal partition parameters.

\begin{theorem}[Optimal Resource Allocation]\label{th:optimal_resource_allocation}
For a dual-mode system operating at $\lambda\in\ldm$ with fixed assignment parameter $p\in\mathcal \Pl$, the delay-optimal server partition parameters are:
\[\alpha^*(p;\lambda) = \dfrac{\frac{\lambda p}{\mlo}\sqrt{\frac{\bar p}{\mlt}} + \lp1-\frac{\lambda \bar p}{\mlt}\rp\sqrt{\frac{p}{\mlo}}}{\sqrt{\frac{p}{\mlo}} + \sqrt{\frac{\bar p}{\mlt}}}\]
and
\[\beta^*(p;\lambda) = \dfrac{\frac{\lambda p}{\mco}\sqrt{\frac{\bar p}{\mct}} + \lp1-\frac{\lambda \bar p}{\mct}\rp\sqrt{\frac{p}{\mco}}}{\sqrt{\frac{p}{\mco}} + \sqrt{\frac{\bar p}{\mct}}}.\]
The corresponding delay under the optimal partitioning is:
\[\tdm^*(p;\lambda) = \dfrac{\lp\sqrt{\frac{p}{\mlo}}+\sqrt{\frac{\bar p}{\mlt}}\rp^2}{1-\frac{\lambda p}{\mlo}-\frac{\lambda \bar p}{\mlt}} + \dfrac{\lp\sqrt{\frac{p}{\mco}}+\sqrt{\frac{\bar p}{\mct}}\rp^2}{1-\frac{\lambda p}{\mco}-\frac{\lambda \bar p}{\mct}}\]
\end{theorem}
\begin{IEEEproof}
The result follows from elementary convex optimization. For fixed $\lambda \in \ldm$ and $p \in \Pl$, the delay function $\tdm(p, \cdot, \cdot; \lambda)$ is jointly convex in $\alpha$ and $\beta$ over the feasible set. Setting partial derivatives with respect to $\alpha$ and $\beta$ to zero yields the unique minimizers - $\alpha^*$ and $\beta^*$.
\end{IEEEproof}
\pad 

\subsection{Interpretation}
We now provide intuition to better interpret the expression for the delay under the optimal partitioning. First, we re-write the expression using the canonical parameters:
\[\tdm^*(p;\lambda) = \dfrac{\lp\sqrt{pf_1}+\sqrt{\bar pf_2}\rp^2}{\mu_0- \lambda \lp pf_1 + \bar pf_2\rp} + \dfrac{\lp\sqrt{p\bar f_1}+\sqrt{\bar p \bar f_2}\rp^2}{K\mu_0-\lambda\lp p\bar f_1 + \bar p \bar f_2\rp}\]
Now, define
\[f(p) \coloneqq pf_1 + (1-p)f_2,\]
which  represents the effective fraction of workload processed locally for a system operating under assignment parameter $p$.

\noindent Now, we can succinctly rewrite the optimal dual-mode delay as:
\begin{equation}\tdm^*(p;\lambda) = \tsm(f(p);\lambda) + \toh(p;\lambda)\label{eqn:intuitive_delay}	\end{equation}

where $\tsm(\cdot;\lambda)$ is the delay function from the tunable-mode system, given by:
\[\tsm(f;\lambda) \coloneqq \frac{f}{\mu_0 - f\lambda} + \frac{\bar{f}}{K\mu_0 - \bar{f}\lambda}\]
and $\toh(\cdot;\lambda)$ denotes a non-negative overhead term:
{\small
\begin{align*}
\toh(p;\lambda)& \\
\coloneqq \:\ &2\sqrt{p\bar p}\lcb \frac{\sqrt{f_1f_2}}{\mu_0- \lambda \lp pf_1 + \bar pf_2\rp} + \frac{\sqrt{\bar f_1 \bar f_2}}{K\mu_0-\lambda\lp p\bar f_1 + \bar p \bar f_2\rp}\rcb
\end{align*}}
which captures the additional delay incurred from allocating server resources across different modes to achieve a desired local-to-cloud workload split -- $f(p)$.

This decomposition shows that the delay of a dual-mode system operating under assignment parameter $p$ can be expressed as the delay of an equivalent tunable-mode system with service fraction $f(p)$, plus an additional overhead cost.

\subsection{System Design Considerations}
The delay decomposition in \eqref{eqn:intuitive_delay} provides a natural lower bound on the dual-mode system's delay:
\begin{proposition}\label{pro:lower_bound}
For any dual-mode system with fixed server capacities $\mu_0$ and $K\mu_0$ (in canonical form), and operating at arrival rate $\lambda \in \ldm$, the delay under any feasible assignment parameter $p \in \Pl$ is lower bounded by the optimal delay achievable in the corresponding tunable-mode system. Formally,
\[\tdm^*(p;\lambda) \geq \tsm^*(\lambda), \quad \forall\, p \in \Pl,\; \lambda \in \ldm.\]
\end{proposition}
\begin{IEEEproof}
The result is immediate from \eqref{eqn:intuitive_delay} and the non-negativity of the overhead term.
\end{IEEEproof}
\pad

Proposition~\ref{pro:lower_bound} indicates that the optimal tunable-mode delay serves as a fundamental performance benchmark, independent of specific mode parameters ($f_1$ and $f_2$) and determined solely by server capacities and arrival rate. 

Having established a lower bound for the delay performance of a dual-mode system, we now examine the conditions under which this bound is achieved. Attaining the lower bound requires that the overhead term in \eqref{eqn:intuitive_delay}, $\toh(p;\lambda)$ vanishes. This occurs for all arrival rates $\lambda$, when the service mode parameters satisfy $f_1 = 0$ and $f_2 = 1$. Equivalently, SM1 processes the job entirely at the cloud and SM2 processes the job entirely at the local server. We establish the optimal assignment strategy for such a system in the following Proposition.

\begin{proposition}\label{pro:attain_lb}
For a dual-mode system, if the canonical service mode parameters satisfy
\[f_1 = 0 \quad \text{and} \quad f_2 = 1\]
then the system attains the delay lower bound from Proposition~\ref{pro:lower_bound} by operating under the assignment parameter
\[p^*(\lambda) = f^*(\lambda),\]
where $f^*(\lambda)$ denotes the delay-optimal service-fraction parameter from the corresponding tunable-mode system.
\end{proposition}
\begin{IEEEproof}
When $f_1 = 0, f_2 = 1$, the optimal server partitionings are trivially $\alpha^* = 0$ and $\beta^* = 1$. Moreover, the effective local processing fraction $f(p) = pf_1 + \bar pf_2 = p$. Thus, setting assignment parameter, $p = f^*(\lambda)$, we have:
\begin{align*}
\tdm^*(f^*(\lambda);\lambda) 
&= \tsm(f^*(\lambda);\lambda) + \toh(f^*(\lambda);\lambda)\\
&= \tsm^*(\lambda)
\end{align*}
\end{IEEEproof}

Proposition~\ref{pro:attain_lb} suggests that the idealized dual-mode configuration that achieves the minimum possible delay corresponds to one mode offloading all processing to the cloud, while the other performs all processing locally. However, such a setup is typically infeasible in practical systems, where jobs are offloaded precisely because local resources are insufficient to handle full workloads. As a result, while instructive as a theoretical benchmark, this configuration lies outside the scope of most real-world offloading scenarios.


\pad
\noindent In the next section, we build on the intuitive delay representation in \eqref{eqn:intuitive_delay} and our tunable-mode analysis to identify a class of systems in which one mode becomes redundant under optimal operation. We also show that in the general case, the optimal strategy exhibits a breakaway structure, which we characterize next.

\section{Optimal Assignment Strategy}\label{sec:optimal_policy}
Having identified optimal server allocations for a given assignment parameter, we now restrict our analysis to systems operating under these optimal partitionings. Our goal in this section is to characterize the optimal assignment parameter $p^*(\lambda)$, defined as:
\[p^*(\lambda) = \min_{p\in\mathcal \Pl} \tdm^*(p;\lambda)\]
which fully specifies the optimal operating point.

Solving for the critical points of $\tdm^*(\cdot;\lambda)$ analytically involves solving an eighth-degree polynomial equation, which generally does not yield a simple closed-form solution and typically requires numerical optimization techniques. However, leveraging insights from our preliminary analysis, we identify scenarios where the optimization problem simplifies significantly.

\subsection{Redundancy of Local-Heavy Mode}
In this subsection, we characterize a family of dual-mode systems in which the "local-heavy" SM2 is redundant and the optimal strategy is to assign all jobs to the "cloud-heavy" SM1. We formalize this result below.

\begin{theorem}[Redundant Mode]\label{th:redundant_mode}
For a dual mode system, if the canonical parameter $f_1$ of Service Mode~1 satisfies:
\[f_1 \geq \frac{1}{K+1}\]
then, the delay optimal assignment parameter,
\[p^*(\lambda) = 1\]
for all $\lambda\in\ldm$.
\end{theorem}
\begin{IEEEproof}
Since $K>1$ and $f_1<f_2$ by convention, the optimal single-mode fraction
$f^*(\lambda)\in\left[0,\frac{1}{K+1}\right)$
for all $\lambda\in\ldm$ and hence, $f^*(\lambda) < f_1 < f_2$
for all $\lambda\in\ldm$.\\
Given the strict convexity of $\tsm(f(p);\lambda)$, it follows that:
\[\tsm(f_1;\lambda) \leq \tsm(f;\lambda) \quad \forall f\in[f_1,f_2]\]

From Theorem~\ref{th:optimal_resource_allocation}, the delay in this system under the optimal server partitionings at assignment parameter $p$ is given by
\begin{align*}
	\tdm^*(p;\lambda) &= \tsm(f(p);\lambda) + \toh(p;\lambda) \\
	&\geq \tsm(f_1;\lambda) + \toh(p;\lambda) \\
	&\geq \tsm(f_1;\lambda) + \toh(1;\lambda) \\
	&=\tdm^*(1;\lambda)
\end{align*}
which establishes the result.
\end{IEEEproof}

Recall that the quantity $\frac{1}{K+1}$ represents the relative capacity of the local server ($\mu_0$) compared to the total system capacity ($(K+1)\mu_0$). Theorem~\ref{th:redundant_mode} highlights an important design insight: if practical constraints cause every mode to assign a local workload fraction exceeding the local server's proportional capacity, then the optimal strategy is to exclusively use the mode that minimizes local processing.


\subsection{Breakaway Behavior of Optimal Assignment}
Having established conditions under which SM2 is redundant, we now consider the complementary case, where \mbox{$f_1<1/(K+1)$}. This includes the class of \textit{throughput-efficient} systems from Definition~\ref{def:throughput_efficient}. We show that the optimal assignment parameter function $p^*(\lambda)$ exhibits a simple breakaway structure.

Specifically, at low loads, it is optimal to exclusively assign jobs and resources to the "cloud-heavy" SM1 (i.e., $p^*(\lambda) = 1$). At higher loads, the optimal policy breaks away from exclusive use of SM1 and begins assigning some jobs to SM2 (i.e., $p^*(\lambda)\in[0,1)$).

We begin by showing the optimal behavior at lower loads in the following theorem.

\begin{theorem}[SM1 Optimality at Low Loads]\label{th:low_arrival}
For a dual mode system, if the parameter $f_1$ of Service Mode~1 satisfies:
\[f_1 < \frac{1}{K+1}\]
Then, the delay optimal assignment parameter,
\[p^*(\lambda) = 1\]
for all $\lambda\in\ldm$ such that \mbox{$\lambda \leq \frac{\mu_0(K-\sqrt K)}{1-f_1(1+\sqrt K)}$.}
\end{theorem}
\begin{IEEEproof}
For all arrival rates $\lambda \leq \frac{\mu_0(K-\sqrt K)}{1-f_1(1+\sqrt K)}$, we have $f^*(\lambda) \leq f_1 < f_2$. The remainder follows from the proof of Theorem~\ref{th:redundant_mode}.
\end{IEEEproof}

Having established that assigning exclusively to SM1 is optimal for lower loads, we now demonstrate that at sufficiently high loads, the optimal policy indeed departs from exclusively assigning SM1. 

\begin{theorem}[Break away from SM1 at High Loads]\label{th:breakaway}
For a dual mode system, if the parameter $f_1$ of Service Mode~1 satisfies:
\[f_1 < \frac{1}{K+1}\]
Then,
the delay optimal assignment parameter,
\[p^*(\lambda) \in [0,1)\]
for all arrival rates, $\lambda\in\ldm$ such that:
\[\lambda \geq \frac{K\mu_0}{1-f_1}\]
\end{theorem}
\begin{IEEEproof} 
The result follows from stability considerations. Full details are omitted for brevity.
%
\end{IEEEproof}

\pad
Therefore, dual-mode systems satisfying $f_1 < 1/(K+1)$ exhibit a clear \textit{breakaway structure} in their optimal assignment policy: initially, the optimal policy assigns exclusively to SM1 at low arrival rates, then transitions toward partial or full assignment to SM2 as system load increases. We characterize the high load behavior of the system in the following subsection.

\subsection{High Load Regime}
In general, the optimal assignment parameter $p^*(\lambda)$ is not necessarily monotonic and may exhibit complex behavior at intermediate loads after breaking away from exclusive SM1 assignment. Since we already characterized the optimal operating strategy for all loads for system with canonical  parameter $f_1>1/(K+1)$, in Theorem~\ref{th:redundant_mode}, in this section, we characterize the behavior of $p^*(\lambda)$ in the high-load regime for systems satisfying \mbox{$f_1<1/(K+1)$}, based on the canonical parameter $f_2$. We distinguish two distinct scenarios: (i) \textit{throughput-efficient} systems with \mbox{$f_2 > 1/(K+1)$} (as defined in Definition~\ref{def:throughput_efficient}), and (ii) systems with \mbox{$f_2 \leq 1/(K+1)$}, where SM2 becomes a throughput bottleneck.

We first establish that, for throughput-efficient systems, the optimal strategy in the high-load regime involves assigning jobs to both service modes:
\begin{theorem}\label{th:needs_both}
For a \textit{throughput-efficient} dual-mode system, the optimal assignment parameter satisfies:
\[p^*(\lambda) \in (0,1)\]
for all $\lambda\in\ldm$ such that:
\[\lambda>\max\lcb\tfrac{\mu_0}{f_2}, \tfrac{K\mu_0}{1-f_1}\rcb\]
Moreover, in the limit of full system utilization, the optimal assignment converges to
\[\lim_{\lambda \rightarrow \lambda_{\max}^-} p^*(\lambda) \to \tfrac{f_2 - 1/(K+1)}{f_2 - f_1}\]
where $\lambda_{\max} = (K+1)\mu_0$.
\end{theorem}
\begin{IEEEproof}
Given $\lambda>\max\lcb\frac{\mu_0}{f_2}, \frac{K\mu_0}{1-f_1}\rcb$, we equivalently have: 
\[\lambda>\mlo \quad \text{and} \quad \lambda > \mct\]
Hence, neither exclusively assigning SM1 nor exclusively assigning SM2 can stabilize the system. Therefore, the optimal policy must utilize both modes, implying $p^*(\lambda) \neq 0,1$.
To establish the limit behavior as $\lambda \to \lambda_{\max}^{-}$, note that in the limit, the set of stabilizing assignment parameters, $\Pl$ converges to the singleton $\lcb\frac{f_2 - 1/(K+1)}{f_2 - f_1}\rcb$. The result follows.
\end{IEEEproof}
Notice that $p = \frac{f_2 - 1/(K+1)}{f_2 - f_1}$ implies $f(p) = \frac{1}{K+1}$ i.e. in the limit of full system utilization, the optimal strategy balances the load between the servers in proportion to their respective capacities.

We now consider the second scenario, establishing that for sufficiently high loads, exclusive assignment to SM2 becomes optimal:
\begin{theorem}\label{th:SM2_opt}
For a dual mode system, if the parameter $f_2$ of Service Mode~1 satisfies:
\[f_2 \leq \frac{1}{K+1}\]
Then, the delay-optimal assignment parameter satisfies:
\[p^*(\lambda) = 0\]
for all arrival rates, $\lambda\in\ldm$ such that:
\[\lambda \geq \frac{\mu_0(K-\sqrt K)}{1-f_2(1+\sqrt K)}\]
\end{theorem}
\begin{IEEEproof}
For arrival rates $\lambda \geq \frac{\mu_0(K-\sqrt K)}{1-f_2(1+\sqrt K)}$, we have $f_1 < f_2 \leq f^*(\lambda)$. The remainder follows from the proof of Theorem~\ref{th:redundant_mode}.
\end{IEEEproof}




\section{Performance Evaluation}\label{sec:simulations}
In this section, we illustrate the analytical insights derived in earlier sections using numerical computation. We focus our investigation on three representative dual-mode systems, each defined by distinct canonical service mode parameters. To facilitate meaningful comparisons, we normalize the local server capacity to $\mu_0 = 1$ and set the cloud-to-local capacity ratio to $K = 4$, resulting in a total system capacity of $\mu^* = (K+1)\mu_0 = 5$. The arrival rate is represented as a normalized load parameter $\rho = \lambda / \mu^*$, signifying the fraction of total system capacity utilized.

The three systems considered are: System A, a throughput-efficient configuration with $f_1 = 0.1$ and $f_2 = 0.3$; System B, a configuration overly reliant on cloud processing, with $f_1 = 0.05$ and $f_2 = 0.15$; and System C, which is overly reliant on local resources, with $f_1 = 0.25$ and $f_2 = 0.4$.

As Systems B and C are not throughput-efficient, they cannot support the full arrival rate up to $\mu^*$. Using Theorem~\ref{th:stability_region}, we determine the maximum stabilizable loads: $\rho_{\max}^A = 1$, $\rho_{\max}^B = 0.94$, and $\rho_{\max}^C = 0.8$.

\begin{figure}[t]
    \centering
    \begin{minipage}{0.47\linewidth}
    	\centering
	    \includegraphics[width=\linewidth]{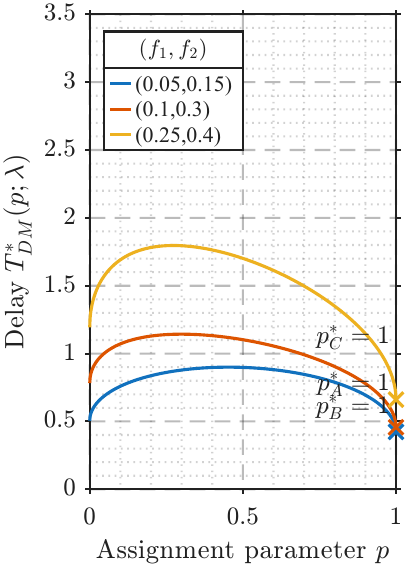}
	    \smallskip
		\small (a) $\rho = 0.3$ ($\lambda=1.5)$
		\smallskip
	\end{minipage}
	\begin{minipage}{0.47\linewidth}
    	\centering
	    \includegraphics[width=\linewidth]{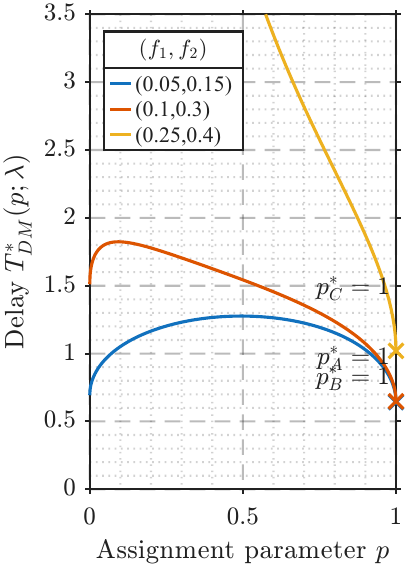}
	    \smallskip
		\small (b) $\rho = 0.5$ ($\lambda=2.5)$
		\smallskip
	\end{minipage}
	\begin{minipage}{0.47\linewidth}
    	\centering
	    \includegraphics[width=\linewidth]{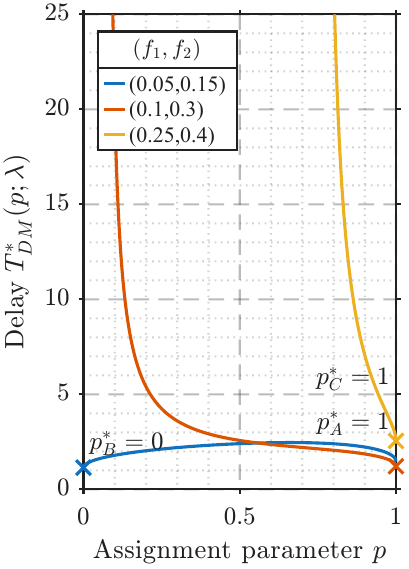}
	    \smallskip
		\small (c) $\rho = 0.7$ ($\lambda=3.5)$
	\end{minipage}
	\begin{minipage}{0.47\linewidth}
    	\centering
	    \includegraphics[width=\linewidth]{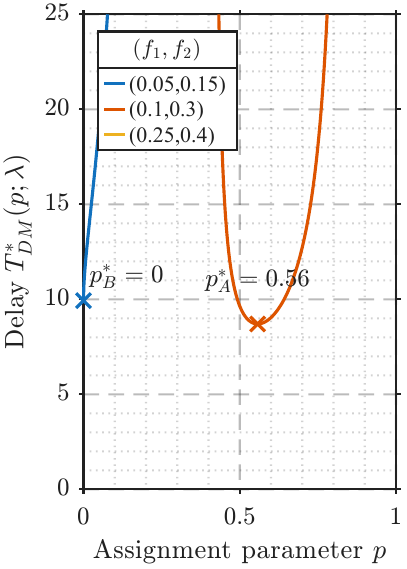}
	    \smallskip
		\small (d) $\rho = 0.92$ ($\lambda=4.6)$
	\end{minipage}
    \caption{Delay $\tdm(p;\lambda)$ vs. assignment parameter $p$ under optimal partitioning, shown for increasing system loads. $p^*$s indicate the optimal assignment parameter.}
    \label{fig:benefit}
\end{figure}


\subsection{Benefit of Optimal Assignment}
In Figure~\ref{fig:benefit}, we plot the delay of each system, denoted by $\tdm^*(p;\lambda)$, as a function of the assignment probability $p$ under optimal server partitioning. Subfigures (a) through (d) represent different load conditions to demonstrate the sensitivity of delay performance to suboptimal assignment choices across feasible values of $p$.

Consistent with our analysis, at low to moderate loads (Figures~\ref{fig:benefit}a-b), all systems achieve optimal delay performance with exclusive SM1 assignment ($p^* = 1$). As loads increase (Figure~\ref{fig:benefit}c), the feasible range of $p$ significantly diminishes in System C, revealing a stability bottleneck due to excessive reliance on local processing. Notably, System B has better delay performance compared to the throughput-efficient System A for a substantial range of $p$ up to a load of about $0.7$.

Approaching system capacity (Figure~\ref{fig:benefit}d), System A achieves better performance under optimal assignment as System B's cloud queues become increasingly congested due to its cloud reliance.

Finally, we observe that optimal selection of the assignment parameter is crucial at higher loads, where small deviations about the optimal operating point result in relatively larger delay penalties.


\subsection{Assignment Dynamics Across Load}
\begin{figure}[!t]
\centering
\includegraphics[width=0.43\textwidth]{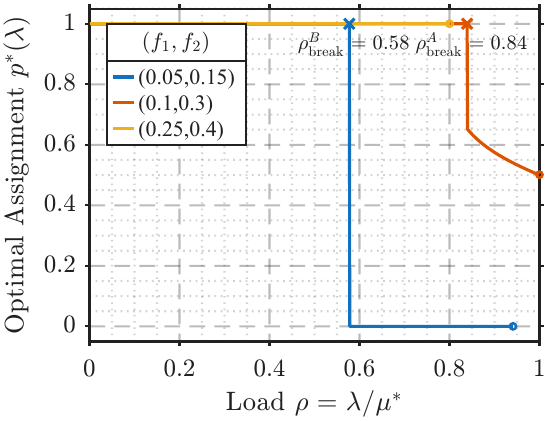}
\caption{Optimal assignment $p^*(\lambda)$ vs. load $\rho$. Systems A and B exhibit breakaway transitions at $\rho_{\text{break}}^A$ and $\rho_{\text{break}}^B$. System C remains at $p^* = 1$.}
\label{fig:opt_p}
\end{figure}

In Figure~\ref{fig:opt_p}, we plot the evolution of the optimal assignment parameter $p^*(\lambda)$ as a function of system load and verify the structural behaviors predicted in our analysis.
Note that the curves for System B and C extend up to the maximum stabilizable load of $0.94$ and $0.8$ respectively.

Consistent with our analysis, System~C which belongs to the family of redundant mode systems described in Theorem~\ref{th:redundant_mode} always operates under exclusive SM1 assignment ($p^* = 1$) across all feasible loads. In contrast, Systems A and B exhibit a breakaway structure, transitioning from exclusive SM1 assignment at lower loads ($p^* = 1$) to mixed or exclusive SM2 assignment at higher loads. We explicitly highlight the breakaway points, $\rho^A_{\text{break}}$ and $\rho^B_{\text{break}}$, demonstrating these load-dependent transitions.

As loads approach system capacity, the optimal assignment in System A converges to $\frac{f_2 - 1/(K+1)}{f_2-f_1} = 0.5$, as indicated by Theorem~\ref{th:needs_both}. Similarly, at higher loads, System B operates under optimal assignment of $p^* = 0$, as established in Theorem~\ref{th:SM2_opt}.

\subsection{Delay under Optimal Assignment}
\begin{figure}[!t]
\centering
\begin{minipage}{0.9\linewidth}
\centering
\includegraphics[width=\textwidth]{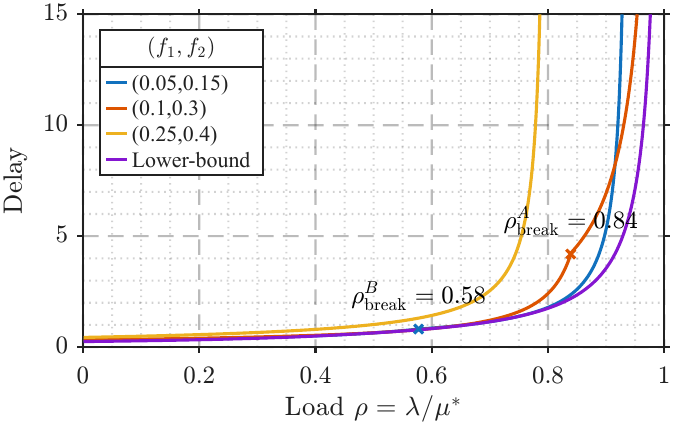}
\smallskip
\small (a) Delay under optimal assignment $p^*(\lambda)$ vs. load $\rho$. Break points mark shifts in assignment strategy.
\bigskip
\end{minipage}
\begin{minipage}{0.9\linewidth}
\centering
\includegraphics[width=\textwidth]{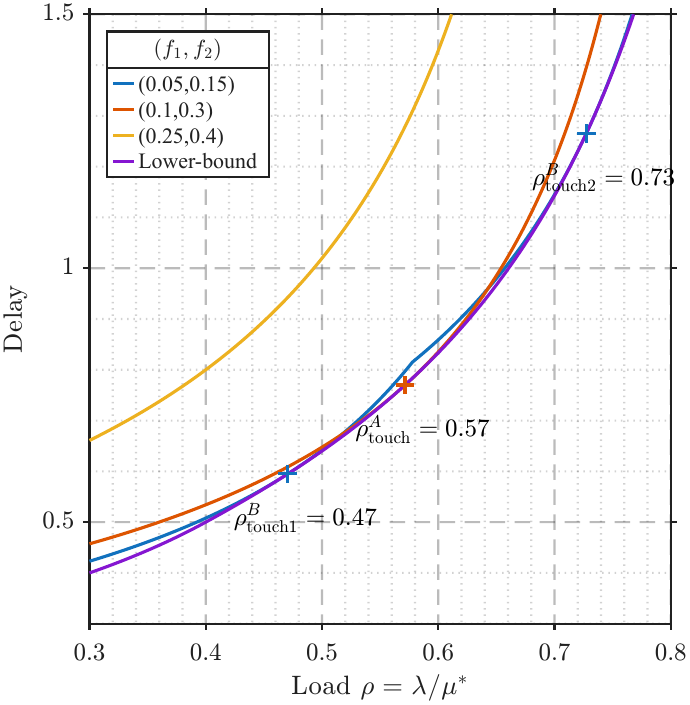}
\smallskip
\small (b) Zoomed-in view highlighting loads where Systems A and B match the tunable-mode lower bound at $\rho_{\text{touch}}^A$, $\rho_{\text{touch1}}^B$, and $\rho_{\text{touch2}}^B$.
\bigskip
\end{minipage}
\caption{Delay performance under optimal assignment compared to the tunable-mode lower bound. Subfigure (b) highlights points where the bound is achieved.}
\label{fig:opt_delay}
\end{figure}

Figure~\ref{fig:opt_delay} compares the delay performance of dual-mode systems operating with the optimal assignment parameter, $\tdm^*(p^*(\lambda);\lambda)$, against system load. We also include the delay performance of a tunable-mode system operating at its optimal service fraction, which serves as a fundamental lower bound for the dual-mode systems (Proposition~\ref{pro:lower_bound}). Figure~\ref{fig:opt_delay}a shows the full plot, while Figure~\ref{fig:opt_delay}b zooms in to highlight key features.

At the breakaway loads ($\rho_{\text{break}}^A, \rho_{\text{break}}^B$), we note a discontinuity in the slope of the delay curve, reflecting changes in assignment strategies.

System~C performs worst across all loads, consistent with our analysis: the tunable-mode optimum is cloud-reliant ($f^*(\lambda) < 1/(K+1)$), whereas both of System~C’s modes are local-heavy ($1/(K+1)<f_1<f_2$).

In Figure~\ref{fig:opt_delay}b, we highlight points where the delay curves for Systems A and B intersect the tunable-mode lower bound at loads $\rho_{\text{touch}}^A = 0.57$, $\rho_{\text{touch1}}^B = 0.47$, and $\rho_{\text{touch2}}^B = 0.73$. As established in our analysis, the optimal service fraction in the tunable-mode system, $f^*(\lambda)$, increases monotonically from 0 at low loads to $1/(K+1)$ as the system approaches full utilization. These intersections occur precisely at the loads where $f^*(\lambda)$ matches the canonical service mode parameters: $f_1$ for System~A, and $f_1$ or $f_2$ for System~B. At these points, the dual-mode system achieves the tunable-mode lower bound by exclusively assigning all jobs to the corresponding service mode. Hence, at lower loads, System~B generally outperforms System~A. However, due to its reliance on cloud-heavy modes, System~B is unable to stabilize near full capacity, leading to substantially higher delays in the high-load regime.

This phenomenon underscores a critical system design trade-off: Throughput-efficient service mode design ($f_1<1/(K+1)<f_2$) allows fully utilizing the servers' capacities and maximizes the stability region. However, systems that are not throughput efficient, with $(f_1<f_2<1/(K+1))$ can outperform the throughput-efficient system at lower loads while sacrificing the ability to stabilize near capacity.

\section{Conclusion}\label{sec:conclusion}
We studied a computation offloading system consisting of a sequential local and cloud server, supporting two distinct service modes differentiated by how processing tasks are divided between these servers. We characterized the system's stability region and derived principles for designing throughput-efficient service modes.

We then derived the optimal server resource allocation for a fixed job assignment strategy and presented a compact expression for delay under these optimal partitionings. This representation allowed us to identify a fundamental lower bound on delay and interpret the dual-mode system by comparison to a tunable single-mode system.

Finally, we showed that under optimal resource allocation, the delay-optimal assignment strategy exhibits a breakaway structure: at lower loads, all jobs are assigned to the cloud-heavy mode, while at higher loads, the system breaks away from this strategy and assigns some or all jobs to the local-heavy mode. Our simulation results confirm and illustrate these theoretical insights.

\appendices
\section{Proofs supporting Theorem~\ref{th:stability_region}}\label{app:stability_region}
\subsection{Proof of Lemma~\ref{lem:acheivability}}\label{app:acheivability}
Given $\lambda \in \ldm$, it follows that $\lambda < \mu^*$.  Algebraic simplifications then yield:
\[p_{\min}<p_{\max}\]
where: $p_{\min} = 1 - \frac{1/\lambda-1/\mlo}{1/\mlt-1/\mlo}$ and $p_{\max} = \frac{1/\lambda-1/\mct}{1/\mco-1/\mct}$.\\

Additionally, since $\lambda < \mlo, \mct$, we have:
\[p_{\min} < 1 \quad \text{and} \quad p_{\max}>0\]

Pick $p^* \in \Pl \coloneqq [0,1]\cap (p_{\min}, p_{\max})$.
Then, $p^*$ is a feasible choice of assignment parameter and satisfies
\begin{align*}
	1 - \dfrac{\frac{1}{\lambda}-\frac{1}{\mlo}}{\frac{1}{\mlt}-\frac{1}{\mlo}} <  p^* < \dfrac{\frac{1}{\lambda}-\frac{1}{\mct}}{\frac{1}{\mco}-\frac{1}{\mct}} 
\end{align*}
Expanding both sides of these inequalities, we get
\begin{align*}
	\frac{\lambda p^*}{\mlo} + \frac{\lambda \bar{p^*}}{\mlt} < 1 \\
	\frac{\lambda p^*}{\mco} + \frac{\lambda \bar{p^*}}{\mct} < 1 
\end{align*}
Once again, since $\lambda<\mu_{l1},\mu_{c2}$, the terms in the LHS of both the above inequalities are non-negative and in $[0,1]$. Define:
\begin{align*}
	\alpha^* = \dfrac{1}{2} \lsb 1 - \frac{\lambda \bar{p^*}}{\mlt} + \frac{\lambda p^*}{\mlo} \rsb \: \text{and} \:\: \beta^* = \dfrac{1}{2} \lsb 1 - \frac{\lambda \bar{p^*}}{\mct} + \frac{\lambda p^*}{\mco} \rsb
\end{align*}
These definitions yield $\alpha^*, \beta^* \in [0,1]$ that satisfy:
\begin{align*}
	\frac{\lambda p^*}{\mlo} < \alpha^* < 1 - \frac{\lambda \bar{p^*}}{\mlt} \quad \text{and} \quad
	\frac{\lambda p^*}{\mco} < \beta^* < 1 - \frac{\lambda \bar{p^*}}{\mct}
\end{align*}
confirming that the inequalities in \eqref{eqn:stability_equations} hold.
\proofover


\subsection{Proof of Lemma~\ref{lem:converse}}\label{app:converse}
To prove by contradiction, suppose $\lambda\notin\ldm$ and the inequalities in \eqref{eqn:stability_equations} are satisfied for some $p,\alpha, \beta \in [0,1]$.

Adding the first two inequalities in \eqref{eqn:stability_equations}, we get
\begin{align}
	\lambda < \alpha \mlo + \bar\alpha \mlt < \mlo(\alpha + \bar\alpha) = \mlo \label{eqn:local_cap}
\end{align}
Similarly, we get 
\begin{align}
	\lambda < \beta \mco + \bar\beta \mlt < \mct \label{eqn:cloud_cap}
\end{align}
Now, remains to show $\lambda < \mu^*$ to show a contradiction. For this, we eliminate $\alpha, \beta$ from \eqref{eqn:stability_equations} to get:
\begin{align*}
	\frac{\lambda p}{\mlo} + \frac{\lambda \bar{p}}{\mlt} < 1 \\
	\frac{\lambda p}{\mco} + \frac{\lambda \bar{p}}{\mct} < 1 
\end{align*}
Eliminating $p$ from these inequalities, we get
\begin{align*}
1 - \dfrac{\frac{1}{\lambda}-\frac{1}{\mlo}}{\frac{1}{\mlt}-\frac{1}{\mlo}} < \dfrac{\frac{1}{\lambda}-\frac{1}{\mct}}{\frac{1}{\mco}-\frac{1}{\mct}}
\end{align*}
which simplifies to 
\begin{align}
\lambda < \mu^* \label{eqn:system_cap}
\end{align}
Now, from \eqref{eqn:local_cap}-\eqref{eqn:system_cap}, we must have $\lambda\in\ldm$ which contradicts our assumption.
\proofover

\section{Proof of Theorem~\ref{th:redundant_mode}}\label{app:redundant_mode}
\begin{IEEEproof}
Since $K>1$ and $f_1<f_2$ by convention, the optimal single-mode fraction
$f^*(\lambda)\in\left[0,\frac{1}{K+1}\right)$
for all $\lambda\in\ldm$. Thus, we have
\[f^*(\lambda) < f_1 < f_2\]
for all $\lambda\in\ldm$.
Given the strict convexity of $\tsm(f(p);\lambda)$ it follows that:
\[\tsm(f_1;\lambda) \leq \tsm(f;\lambda) \quad \forall f\in[f_1,f_2]\]

From Theorem~\ref{th:optimal_resource_allocation}, the delay in this system under the optimal server partitionings at assignment parameter $p$ is given by
\begin{align*}
	\tdm^*(p;\lambda) &= \tsm(f(p);\lambda) + \toh(p;\lambda) \\
	&\geq \tsm(f_1;\lambda) + \toh(p;\lambda) \\
	&\geq \tsm(f_1;\lambda) + \toh(1;\lambda) \\
	&=\tdm^*(1;\lambda)
\end{align*}
which establishes the result.
\end{IEEEproof}

\bibliographystyle{IEEEtran}
\bibliography{references.bib}
\end{document}